\documentclass[11pt]{article}
\usepackage{ifthen,amsmath,amssymb,amsthm,latexsym,color,epsfig,bm,verbatim,hyperref}
\usepackage{multirow}

\usepackage{pstricks}
\usepackage{pst-tree}
\usepackage{pst-node}
\usepackage{pst-plot}
\usepackage{wrapfig}

\newtheorem{theorem}{Theorem}[section]

\theoremstyle{definition}

\makeatletter
\@addtoreset{claim}{theorem}
\makeatother

\usepackage{pstricks}
\usepackage{pst-tree}
\usepackage{pst-node}

\psset{unit=0.9pt}

\begin{document}

\title{A Proof of Levi's Extension Lemma}

\author{
{Marcus Schaefer
} \\
{\small School of Computing} \\[-0.13cm]
{\small DePaul University} \\[-0.13cm]
{\small Chicago, Illinois 60604, USA} \\[-0.13cm]
{\small \tt mschaefer@cdm.depaul.edu}\\[-0.13cm]
}

\maketitle

\begin{abstract}
 We give a short and self-contained proof of Levi's Extension Lemma for pseudoline arrangements.
\end{abstract}

\section{Levi's Extension Lemma}

A {\em pseudoline arrangement} is a collection of simple, closed curves (the {\em pseudolines}) in the projective plane so that every two curves cross exactly once. In 1926 Friedrich Levi published the following fundamental extension result on pseudoline arrangements, now named after him.

\begin{theorem}[Levi's (Extension/Enlargement) Lemma]\label{thm:L}
  Given a pseudoline arrangement and two points $p$ and $q$ not lying on the same pseudoline, a pseudoline through $p$ and $q$ can be added to the arrangement.
\end{theorem}

We give what we believe to be a short and self-contained proof of this result. In the next section, we review previous proofs of Levi's Lemma. A {\em bigon} consists of two arcs which intersect in their endpoints and bound a region homeomorphic to a disk.

\begin{proof}
 If $p$ lies on a pseudoline $\alpha$, we can draw a pseudoline $\gamma$ by following $\alpha$ closely on one side, starting and ending at $p$, and so that $\gamma$ crosses $\alpha$, and all other pseudolines containing $p$, in $p$. 
 \marginpar{
\includegraphics[angle=00,origin=c]{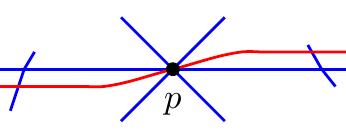}
}
 If $p$ does not lie on a pseudoline, we add a pseudoline as just described, that crosses the boundary of a face of the arrangement that contains $p$. The pseudoline can then be deformed slightly within the face to pass through $p$. In either case, we obtain a new pseudoline arrangement, with a pseudoline $\gamma$ passing through $p$.

 We want to redraw $\gamma$ so it passes through $q$ and remains a pseudoline. If $q$ does not lie on a pseudoline, we pick a curve $\gamma_q$ connecting $\gamma$ with $q$ so that $\gamma_q$ does not cross $\gamma$, and it crosses all curves finitely often. Reroute $\gamma$ sot it closely follows $\gamma_q$ and passes through $q$. 
  \marginpar{
\includegraphics[angle=00,origin=c]{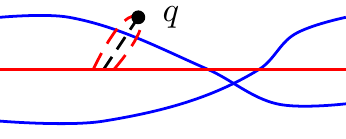}
}
 While $\gamma$ may not be a pseudoline anymore, it is still a simple, closed curve, and it crosses all other curves an odd number of times. If $\gamma$ crosses a pseudoline $\alpha$ more than once, it crosses it at least three times. Follow $\gamma$ as it crosses $\alpha$. Two consecutive crossings along $\gamma$ create an arc which forms a bigon with an arc of $\alpha$. Since there are at least three crossings, we have three such arcs, so one of them, call it $\gamma'$, contains neither $p$ nor $q$, and forms a bigon with an arc $\alpha' \subseteq \alpha$.
 Any pseudoline passing through $\alpha'$ must also pass through $\gamma'$, since $\alpha'$ and $\gamma'$ bound a disk, and a pseudoline can cross $\alpha$, and thus $\alpha'$, at most once. This allows us to detour $\gamma$ so that instead of passing through $\gamma'$ it closely follows $\alpha'$ without crossing it, reducing the number of crossings between $\gamma$ and $\alpha$, and not introducing any new crossings. 
  \marginpar{
\includegraphics[angle=00,origin=c]{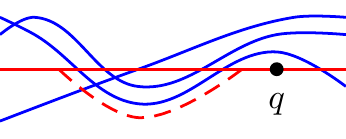}
}
 We repeat this, until $\gamma$ crosses all curves at most once, and has thereby become part of the pseudoline arrangement.

 If $q$ lies on a pseudoline, we proceed similarly. Suppose $q$ lies on pseudoline $\beta$. Since $\gamma$ crosses $\beta$ once, we can choose $\gamma_q$ connecting $\gamma$ to $q$ as a subarc for $\beta$ connecting the crossing to $q$. As earlier, we reroute $\gamma$ along $\gamma_q$ and pass it through $q$. If $q$ is the crossing of several pseudolines, we can route $\gamma$ at $q$ so it crosses every pseudoline crossing $q$ exactly once in $q$.
 All pseudolines still cross $\gamma$ an odd number of times. Suppose some pseudoline $\alpha$ crosses
 $\gamma$ more than once. If $\alpha$ contains neither $p$ nor $q$ we can remove all but one crossing of $\gamma$ with $\alpha$ as we did earlier. This leaves us with the case that $\alpha$ contains either $p$ or $q$ (not both), let us assume $p$. If $\gamma$ and $\alpha$ cross at least five times, then there must be an arc along $\gamma$ which is not incident on $p$, and does not contain $q$. We can then remove two crossings between $\gamma$ and $\alpha$ as earlier, by rerouting $\gamma$ along $\alpha$. Therefore $\gamma$ and $\alpha$ cross exactly three times, and one of those crossings is $p$. If $q$ lies on one of the two $\gamma$-arcs incident to $p$, we can reroute the third $\gamma$-arc along $\alpha$ as before. Hence $q$ lies on that third arc. The two $\gamma$-arcs incident to $p$ cannot overlap on $\alpha$ (this would require additional crossings for $\gamma$ to escape from a bigon), which means we can reroute both arcs simultaneously, while keeping the crossing in $\gamma$ with $\alpha$ in $p$. 
  \marginpar{
\includegraphics[angle=00,origin=c]{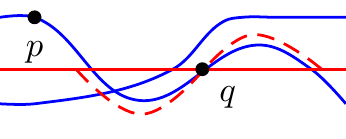}
}
\end{proof}

\section{Brief History}

Levi proved the extension lemma as property $(3)$ of pseudoline arrangements in the paper which formally introduced pseudoline arrangements into the literature~\cite{L26}. His goal was to show that certain results on line arrangements do not depend on the geometry of these arrangements, but on their topology. The result has become known as Levi's Lemma (a name also applied to a result on semigroups also due to Friedrich Levi), or Levi's Enlargement/Extension Lemma. The result is cited often in the literature, but proofs are relatively rare.\footnote{Zentralblatt lists 39 citations of Levi's paper, google scholar 125; not all of these necessarily refer to the lemma, of course. MathSciNet does not list the paper at this point.}

Levi's original proof is elegant and short, and arguably, quite similar to the proof presented here (Levi uses a different argument for treating the case that both $p$ and $q$ lie on multiple pseudolines). Its main disadvantage is that it is written in German. Gr\"{u}nnbaum published the first English proof in his book {\em Arrangements and Spreads}~\cite{G72c}; he uses ideas from Levi's original proof, but uses a different approach.

In 1991, Snoeyink and Hershberger otbained a generalization of Levi's lemma (allowing curves to cross up to twice) using a sweeping argument~\cite{SH91}. Sturmfels and Ziegler, in a 1993 paper, gave a proof of Levi's Lemma using the machinery of oriented matroids~\cite{SZ93}. Felsner and Weil give a very short proof of Levi's lemma as a consequence of a sweeping procedure for pseudoline arrangements~\cite{FW01}. Arroyo, McQuillan, Richter, and Salazar~\cite{AMcQRS18} finally present a proof which is very close in spirit to Levi's original proof, but emphasizes the dual face structure more than Levi did.

The proof given in this short note differs (slightly) from previous proofs by using a redrawing argument rather than a drawing argument. For moving points $a$ and $b$, the proof contains an algorithm on how to update the pseudoline correspondingly.

\bibliographystyle{plain}
\bibliography{Levi}

\end{document}